\documentclass[12pt]{article} 

\newcommand*{\myEXPfont}{\fontfamily{cmr}\selectfont}
\usepackage{graphicx} 
\usepackage{mathtools}
\usepackage{amsmath}
\usepackage{multirow}
\usepackage[export]{adjustbox}
\usepackage[left=1in, right=1in, top=1in, bottom=1in]{geometry} 
\usepackage{booktabs}
\usepackage{rotating}
\usepackage{rotfloat}
\usepackage{float}
\usepackage{subfig}
\usepackage{url}
\usepackage{pdfpages}
\usepackage{rotating}
\usepackage[doublespacing]{setspace} 
\usepackage{csvsimple}
\usepackage{array}
\usepackage{wrapfig}
\usepackage[utf8]{inputenc}
\usepackage{tabularx}
\usepackage{float}

\usepackage{amsthm}                
\usepackage{amssymb}

\usepackage{multirow}
\usepackage{lscape}
\usepackage{longtable}

\usepackage{caption}
\newtheorem{thm}{Theorem}

\newtheorem{prop}[thm]{Proposition}

\usepackage{amsfonts}
\usepackage{amsmath} 
\usepackage{tikz}
\usetikzlibrary{positioning,shapes.geometric}
\usepackage{comment}
\usepackage{xcolor}

\usepackage{setspace}

\DeclareMathOperator{\E}{\mbox{{\myEXPfont \textup{E}}}}

\usepackage [english]{babel}
\usepackage [autostyle, english = american]{csquotes}
\MakeOuterQuote{"}

\usepackage{mathrsfs}

\makeatletter
\newcommand*{\indep}{%
  \mathbin{%
    \mathpalette{\@indep}{}%
  }%
}
\newcommand*{\nindep}{%
  \mathbin{
    \mathpalette{\@indep}{\not}
  }%
}
\newcommand*{\@indep}[2]{%
  \sbox0{$#1\perp\m@th$}
  \sbox2{$#1=$}
  \sbox4{$#1\vcenter{}$}
  \rlap{\copy0}
  \dimen@=\dimexpr\ht2-\ht4-.2pt\relax
  \kern\dimen@
  {#2}%
  \kern\dimen@
  \copy0 
} 
\makeatother

\usepackage{xr-hyper}

\definecolor{forestgreen}{RGB}{34,139,34}
\usepackage{hyperref}
\hypersetup{
    colorlinks=true,
    linkcolor=black,
    filecolor=black, 
    citecolor=forestgreen,      
    urlcolor=blue
}

\usepackage{sectsty}
\sectionfont{\large}
\renewcommand{\refname}{REFERENCES}

\usepackage{layout}

\usepackage{geometry}

\newcolumntype{C}[1]{>{\centering\arraybackslash}p{#1}}

\usepackage{authblk}

\usepackage{tabularray}
\usepackage{colortbl}

\usepackage{afterpage}
\usepackage{lscape}
\usepackage{graphicx}

\usepackage{datetime}

\makeatletter
\newcommand*{\addFileDependency}[1]{
  \typeout{(#1)}
  \@addtofilelist{#1}
  \IfFileExists{#1}{}{\typeout{No file #1.}}
}
\makeatother

\usepackage[
    backend=biber,
    style=numeric-comp, 
    sorting=none,       
    giveninits=true,    
    terseinits=true     
]{biblatex}

\DeclareFieldFormat{labelnumberwidth}{[#1]}
\DeclareFieldFormat{shorthandwidth}{[#1]}

\usepackage{etoolbox}
\patchcmd{\endproof}
  {\popQED}
  {\par\popQED}
  {}
  {}

\usepackage{titlesec}
\titlespacing*{\subparagraph}{0pt}{3.25ex plus 1ex minus .2ex}{1em}

\begin{document}

\title{Identification strategies for combining an experimental study with external data}

\author[1,2]{Lawson Ung}
\author[1,3]{Guanbo Wang}
\author[4]{Sebastien Haneuse}
\author[1,2,4]{Miguel A. Hern\'an}
\author[1,2,4,5]{Issa J. Dahabreh}

\affil[1]{CAUSALab, Harvard T.H. Chan School of Public Health, Boston, MA}
\affil[2]{Department of Epidemiology, Harvard T.H. Chan School of Public Health, Boston, MA}
\affil[3]{The Dartmouth Institute for Health Policy and Clinical Practice, Dartmouth Geisel School of Medicine, Lebanon, NH}
\affil[4]{Department of Biostatistics, Harvard T.H. Chan School of Public Health, Boston, MA}
\affil[5]{Richard A. and Susan F. Smith Center for Outcomes Research, Beth Israel Deaconess Medical Center, Boston, MA}


\maketitle{}
\thispagestyle{empty} 



\noindent \textbf{Address for correspondence:} Dr. Lawson Ung, Department of Epidemiology, Harvard T.H. Chan School of Public Health, Boston, MA 02115; email: \href{mailto:lawson_ung@hsph.harvard.edu}{lawson\_ung@hsph.harvard.edu}; phone: +1 (617) 495‑1000.

\thispagestyle{empty}

\clearpage
\thispagestyle{empty} 
\vspace*{0in}
\begin{abstract}
\noindent
\linespread{1.3}\selectfont

\noindent There is increasing interest in combining information from experimental studies, including randomized and single-group trials, with information from external experimental or observational data sources. Such efforts are usually motivated by the desire to compare treatments evaluated in different studies -- for instance, by constructing external comparator groups for some index study -- or to estimate treatment effects with greater precision. Proposals to combine experimental studies with external data were made at least as early as the 1970s, but in recent years have come under increasing consideration within clinical practice, academia, and by regulatory agencies involved in drug and device evaluation, particularly with the increasing availability of trial and observational data. In this paper, we describe basic study templates that combine information from experimental studies with external data, and use the potential (counterfactual) outcomes framework to elaborate identification strategies for potential outcome means and average treatment effects. We argue that these identification
strategies inherit ideas relevant to the study of causation in single-source studies
and the related literature on combining information (e.g., generalizability and transportability methods), but merit consideration as a separate class of causal problems because they differ in terms of their scientific motivations, definitions of the target population, sampling, data structures, and identifiability conditions. In formalizing identification strategies for the analyses described herein, we hope to provide a conceptual foundation to support the systematic use and evaluation of such efforts. \end{abstract}


\clearpage
\section{INTRODUCTION}

Some studies combine information from experiments, including randomized trials and single-group trials, with information from external data sources to answer causal questions \cite{dahabreh2024combining}. Originally proposed in the 1970s in the context of cancer therapy \cite{gehanfreireich1974, pocock1976combination, pocock1979allocation, fleming1982}, such study designs have been described as combining trials with ``historical'' \cite{sacks1982, dempster1983combining, krisam2020enhancing, suissasat2020, ghadessi2020roadmap, schuler2022increasing, liao2023prognostic}, ``external'' \cite{fda2023, burcurwd2020, thorlundextcont2020}, or ``hybrid'' \cite{wang2024evaluating, segalhybrid2020, ventz2022} controls, ``external comparator arms'' \cite{rippinreview2022, tan2022}, or viewed as examples of so-called ``data fusion'' \cite{bareinboim2016causalfusion, breskin2021fusion, shook2024fusing}. Naturally, interest in conducting such studies has coincided with growing access to individual-level data from trials, and real-world data derived from observational sources, such as electronic health records, healthcare claims, and registries.

Approaches for combining information from experimental studies with external data have been motivated by the lack of direct (head-to-head) treatment comparisons in trials, or the desire to ``borrow strength'' from non-trial participants to achieve more precise effect estimation \cite{carrigan2019, pocock1976combination, ventz2022}. Developing methods to accomplish these goals could support clinical decision-making; provide information for regulatory review \cite{ema2006, hatswell2016regulatory, limminimizing2018, hall2021, curtis2023regulatory, fda2023, wang2023current, ema2025external}; and inform the design and conduct of future clinical trials, for instance by offering provisional evidence for interventions that warrant further study, determining reasonable treatment allocation strategies, and possibly reducing sample size requirements \cite{gray2020, rahman2021, karlsson2024robust}. However, the strong and typically unverifiable assumptions required for such analyses to be endowed with a causal interpretation are not always clearly communicated in applications, guidance documents, nor the statistical literature. 

Here, we describe identification strategies that combine information from experimental studies with external data to learn about causal parameters in a well-defined target population of interest, across a set of basic study templates and data configurations. We use the potential (counterfactual) outcomes framework \cite{splawaneyman1990, rubin1974, robins1986} to organize and extend ideas in the selected literature \cite{gehanfreireich1974, pocock1976combination, pocock1979allocation, sacks1982, breskin2021fusion, ghassami2022combining, ventz2022, li2023improving, valancius2024causal, shook2024fusing}, and identification approaches that involve explicit intervention on participation within the experimental study and of treatment assignment itself. Our systematization of these identification strategies addresses concerns relevant to the study of causation within single-source analyses \cite{rubin1974, robins1986}, as well as causal methods for combining information -- for instance, the presence of study engagement effects \cite{dahabreh2021studydesigns, dahabreh2020transportingStatMed} and the extent to which the study populations are exchangeable. However, differences in the scientific motivation, target population, sampling, data structures, and identifiability assumptions suggest that these analyses merit consideration as a distinct class of causal problems. Our proposed approach to formalizing identification strategies when combining information in the manner described may provide conceptual clarity to support the systematic conduct and evaluation of such work.

\section{SETUP AND DATA} \label{section_setup}

\subsection{Sampling scheme} We consider an experimental study, either a randomized trial (comparing two or more treatments) or a single-group trial (with a single treatment), which we term the \textit{index trial}. We work within a superpopulation framework \cite{robins1988confidence} and a non-nested trial design, which describes settings where trial data and the external data are sampled separately from their respective underlying populations, with sampling fractions not known to the investigators \cite{dahabreh2020transportingStatMed, dahabreh2021studydesigns}. This non-nested trial sampling scheme is by far the most common in applications -- typically, the planning and conduct of the trial occurs independently from the collection of external data that can be drawn from separately conducted studies (experimental or non-experimental), registries, or routinely collected data sources, such as electronic health record or healthcare claims databases. Under the assumptions of non-nested sampling, the combination of the index and external data does not correspond to any well-defined population, thereby constraining identification of causal effects to the populations underlying either the index or external data. That said, our approach can be extended to nested trial designs where the experimental study is embedded within a cohort sampled from the target population of interest. For further details on study designs and sampling schemes, see \cite{dahabreh2020transportingStatMed, dahabreh2021studydesigns}.

\subsection{Notation} We use $X$ to denote baseline (pre-treatment) covariates, $S$ to denote trial participation ($S = 1$ for participants in the index trial; $S=0$ for non-participants), $A$ to denote a discrete treatment, and $Y$ to denote the outcome measured at the end of the study (e.g., binary, count, or continuous). We use $\mathcal A_{S=s}$ to denote the set of treatments in use in the population with $S = s$. Upper-case letters denote random variables and corresponding lower-case letters denote realizations, and $f(\cdot)$ is generic notation for densities. 

\subsection{Data} We work with a composite dataset formed by appending the trial data with a sample from the population underlying the external data \cite{dahabreh2020transportingStatMed}. Within each source $S \in \{0,1\}$, the observations in the composite dataset are independent and identically distributed realizations of the random tuple $(X, S=s , A, Y)$. The distributions underlying the two sources, however, need not be the same. For example, the case-mix (i.e., the marginal distribution of $X$) or the treatment assignment mechanism (i.e., the conditional distribution of $A$ given $X$) may vary across populations. Of note, we allow the set of treatments adopted within the population underlying the index trial to be different from the set of treatments adopted in the population underlying the external data. In other words, we do not require $\mathcal A_{S=1}$ to be the same as $\mathcal A_{S=0}$, implying that the conditional densities $f(a|x,S=s)$ may be 0 for some values of $a$. Probabilities and expectations are defined with respect to the distribution induced by forming the composite dataset. In the following sections, we discuss and present data structures in the context of specific identification strategies relevant to the different treatments that may be adopted in the populations underlying the composite dataset. 

\subsection{Simplifying assumptions} 

For simplicity, we assume complete adherence to the assigned or recommended treatment \cite{dahabreh2022nonadherence}, no loss to follow-up, or competing risks. Our results could be extended to accommodate these issues using methods well-understood in the context of experimental or observational studies \cite{hernan2017perprotocol}. Here, we focus on basic settings to highlight key conceptual issues and the strength of assumptions required to combine information from an experimental study with data from an external source, even before the introduction of aforementioned complexities that arise in applied work. 

\section{CAUSAL ESTIMANDS} \label{section_estimands}

To define causal estimands, we use potential (counterfactual) outcomes \cite{rubin1974, robins1986, splawaneyman1990, robins2000d}. Let $A^{s}$ be the counterfactual treatment under intervention to set trial participation $S$ to $s$, where $S \in \{0,1\}$ and $\mathcal{A}_{S=1} \cup \mathcal{A}_{S=0} \subseteq \{0, 1, 2\}$; and $Y^{s,a}$ be the counterfactual outcome under joint intervention to set trial participation $S$ to $s$ and treatment $A$ to $a$. Under non-nested sampling, the identification strategies we consider are relevant for comparing features of the distribution of potential outcomes under different interventions in the population underlying the index trial ($S = 1$). Here, we focus on potential outcome means under joint intervention to set trial participation $S$ to $s=1$ and treatment $A$ to $a$, $\E[Y^{s=1,a}|S = 1]$, as well as the average treatment effect comparing two treatments $a$ vs. $a'$ on the causal risk difference scale, that is $\E[Y^{s=1,a}-Y^{s=1,a'}|S = 1]$. Identifiability conditions introduced in the next section will permit omission of the $s=1$ counterfactual superscript where appropriate, such that $\E[Y^{s=1,a}|S=1]=\E[Y^a|S=1]$. However, we prefer to be explicit about intervening on $S$ because participation in a trial is typically accompanied by actions \textit{beyond} the assigned treatment itself, and their effects (or lack thereof) on the outcome need to be explicitly reasoned about for identification to proceed \cite{ung2025generalizing}. Finally, we note that our target population for this estimand is the one for which trial participants can be viewed as a simple random sample. Informally, the flow of information in our analyses -- from the population underlying the external data to the population underlying the trial -- is the ``reverse'' of most transportability studies where information flows from the trial to some other, usually broader, target population \cite{westreich2017, rudolph2017, dahabreh2020transportingStatMed}.  

\section{CORE IDENTIFIABILITY CONDITIONS} \label{assumptions}

\subsection{Absence of study engagement effects}

A central assumption underlying our work is the absence of study, or trial, engagement effects \cite{dahabreh2019commentaryonweiss, dahabreh2020transportingStatMed, dahabreh2022nonadherence}. These are the effects of activities or actions related to engagement in a specific study that may affect the outcome through pathways that do not involve the assigned (and, under complete adherence, received) treatment. The assumption that study engagement effects are absent is more likely to hold when participation in a study does not directly affect the behavior of participants in ways that are relevant to the outcome of interest. This is a strong assumption that may preclude, for example, the presence of Hawthorne effects \cite{landsberger1958} and the provision of ancillary medical care or follow-up procedures offered in the trial but not in the population underlying the external data \cite{dahabreh2020benchmarking, dahabreh2020toward}. When study engagement effects are assumed to be absent, investigators may identify the effect of treatment $A$ without regard to the specific setting in which the treatment was administered.

More formally, the assumption that study engagement effects are absent means that for each observation $i$, for each $s \in \{0,1\}$, and every $a \in \mathcal A_{S=1} \cup \mathcal A_{S=0}$,  if $A_i = a$, then $Y_i^{s,a} = Y_i^a$ \cite{dahabreh2022nonadherence, dahabreh2019identification}. The absence of study engagement effects implies, for example, that $\E [Y^{s,a} | S = 1] = \E [Y^{a} | S = 1]$ for all $a$ and $s$. We note that this condition reflects an exclusion restriction such that on a causal directed acyclic graph \cite{robins1986, hernan2024} there would be no directed paths from the node $S$ to the node $Y$ that do not intersect the node $A$. This technical description formalizes the clinical gestalt, often reported in guidance documents on the use of external data \cite{ema2006, fda2023}, that one should exercise caution when combining information from populations that may differ with respect to background standards of care. Since the absence of study engagement effects is often implausible in applied settings, we address how identification can be salvaged when such effects are present in Section 5.5. 

\subsection{Consistency, exchangeability, and positivity} 

The conditions below are used in multiple results presented in the following sections. These conditions are given in terms of $a$, a generic value to which treatment $A$ can be set. When we invoke these conditions in results presented below, the conditions will be applied to specific realizations  of treatment $A$, as appropriate for each data structure under consideration. 

\vspace{0.3in}
\noindent (A1) \textit {Consistency of potential outcomes with respect to treatment $A$.} For each individual $i$ and for each $a \in \mathcal A_{S=1} \cup \mathcal A_{S=0}$, if $A_i = a$, then $Y_i^a = Y_i$. That is, if the $i^{th}$ individual receives treatment $A = a$, then their potential outcome under intervention $a$ is equal to their observed outcome. For consistency to hold, the treatments and outcomes of interest must be well-defined, and concordant between the index trial and the external data \cite{rubin1980randomization, rubin1986, greenland1999, hernan2011compound}. A common violation of the consistency assumption is when treatment initiation practices -- for example, the dose, route, frequency of administration, concomitant interventions, and use of blinding -- differs between the index and external data in an outcome-relevant manner. Another common violation is the failure to reason explicitly about differences in the meaning of treatment assignment in a trial compared to external observational data \cite{dahabreh2025assignment}, where treatment assignment is rarely, if ever, randomized. Such discrepancies would need to be explained away for one to represent these notions of treatment using the same counterfactual random variable $A$ \cite{vanderweele2009further}. 

\vspace{0.3in}
\noindent (A2) \textit {Conditional mean exchangeability over treatment $A$ in the population underlying the index trial.} For every covariate pattern $x$ with positive density in the trial population $f(x,S = 1) > 0$ and for treatment $a \in \mathcal A_{S=1}$, $\E[Y^a |X = x , S = 1 ] = \E[Y^a | X = x, S = 1 , A = a]$. This condition implies that, for every covariate pattern $x$ that exists in the population underlying the index trial, for treatments $a,a' \in \mathcal A_{S=1}$, $$\E[Y^a - Y^{a^\prime} | X = x , S = 1 ] = \E[Y^a | X = x, S = 1 , A = a] - \E[Y^{a^\prime} | X = x, S = 1 , A = a^\prime].$$ Condition (A2) is implied by stronger distributional assumptions. These include $(Y^a, X) \indep A | S = 1$, which is assumed to hold by design when the index trial is marginally randomized; and $Y^a \indep A | (X , S = 1)$, which is assumed to hold by design when the index trial is conditionally or marginally randomized.

\vspace{0.3in}
\noindent (A3) \textit {Positivity of treatment assignment in the index trial.} For every covariate pattern $x$ with positive density in the trial $f(x, S=1) > 0$ and for treatment $a \in \mathcal A_{S=1}$, $\Pr[A = a | X = x, S = 1] > 0$. That is, in the index trial, there is a non-zero probability of being assigned to the treatment levels of interest, conditional on the covariates $X$ required for conditional exchangeability over treatment $A$. This condition holds by design in randomized trials. 

\vspace{0.3in}
\noindent (A4) \textit {Conditional mean exchangeability (transportability) over trial participation $S$.} For every covariate pattern $x$ with positive density $f(x, S=1) > 0$ and for treatment $a \in \mathcal A_{S=0}$, $\E[Y^a|X = x, S = 1] = \E[Y^a|X = x, S = 0]$. This condition, which we  henceforth refer to as \textit{transportability in mean}, is characteristic of generalizability or transportability methods \cite{pearl2011, bareinboim2012transportability, dahabreh2021studydesigns}, and is distinct from exchangeability over treatment (i.e., claims of no unmeasured treatment-outcome confounding). This is because the former involves the assumption that some features of the distribution of potential outcomes under intervention on treatment $A$ are independent of trial participation status $S$, rather than of treatment $A$ itself \cite{pearl2011}. Informally, this assumption requires that the populations underlying the index and external data are comparable within levels of measured covariates, possibly with respect to treatment intervention levels that were in fact not observed across both populations. 

\vspace{0.3in}
\noindent (A5) \textit{Positivity for the population underlying the external data.} For every covariate $x$ with positive density in the population underlying the index trial, the probability of participation in the external data is positive: if $f(x, S = 1) > 0$, then $\Pr[ S = 0 | X = x ] > 0$. That is, the distribution of covariate patterns $x$ needed for conditional mean exchangeability must have common support between the populations underlying the index trial and the external data. 

\subsection {Reasoning about positivity of trial participation} 

In data structures that combine an experimental study with external data for the purposes described in this paper, one assumes that the covariate patterns that can occur in the index trial, required for relevant transportability assumptions to hold, overlap with those that can occur in the population underlying the external data. This condition may be reasonable if external data are derived from sources that are broadly representative of clinical practice. On the other hand, in conventional transportability analyses one usually requires that the covariate patterns that occur in the nonrandomized target population, required for conditional exchangeability over trial participation $S$, are also those that can occur in the index trial \cite{dahabreh2019generalizingbiometrics, dahabreh2020transportingStatMed}. Thus, in transportability applications, investigators may be more likely to encounter positivity violations, for instance when certain subgroups within the target population have not been represented in the index trial owing to strict eligibility criteria \cite{dahabreh2020transportingStatMed}. The asymmetry in these positivity conditions arises because, as emphasized earlier, the target population of interest for the methods presented herein is the population underlying the trial.

\section{IDENTIFICATION} \label{identification}

We now present identification results for various causal estimands specific to basic study templates that combine experimental studies with external data. We first provide well-known identification results using information only from the index trial. Thereafter, we organize identification results organized by the distribution of treatment in the population underlying the external data. More specifically, in the external data, we consider settings where there is a single treatment that is \textit{uniformly adopted by all individuals} (i.e., all individuals have the same treatment value), and also when there is \textit{treatment variation} (i.e., treatment may take different values). We conclude by considering two methodological extensions: salvaging identification in the presence of trial engagement effects, and when the index and external data are combined to learn about some third target population of interest.

We define some statistical estimands using dual subscripts for realizations of trial
participation S and treatment A in that order. For example, we define the functional $$\gamma_{s,a}\equiv \int\E[Y|X=x,S=s,A=a]dF(x|S=1) \equiv \E \big[ \E [Y|X, S=s, A = a] \big| S = 1 \big],$$
where $F(x|S=1)$ is the cumulative distribution function of baseline covariates in the population underlying the index trial. For brevity, in the main text, we only present g-formula identification results \cite{robins1986}. Complete proofs, including weighting re-expressions of identification results based on the g-formula, can be found in the Supplement.

\subsection{Identification using conditions only in the index trial}

First, we provide a well-known identification result for the potential outcome mean under intervention to set treatment $A$ to $a=1$, in the population underlying the index trial. 

\begin{prop}\label{prop1}
Under conditions A1, A2, and A3 for $a=1$, the potential outcome mean under intervention to set treatment $A$ to $a=1$ in the population underlying the index trial, $\E[ Y^{a=1}| S = 1]$, is identified with $$\gamma_{1,1} \equiv \E \big[ \E [Y|X, S=1, A = 1] \big| S = 1 \big].$$
\end{prop}

The identification result for the potential outcome mean under intervention to set treatment $A$ to $a=1$ is the conditional outcome mean among individuals receiving the experimental treatment $A=1$ in the population underlying the index trial, marginalized to its covariate distribution. We use this result in all subsequent identification results concerning average treatment effects because the index trial is the only source of observed information for the potential outcome mean when treatment $A$ is set to $a=1$.  

\subsection{Identification under uniform use of a single treatment in the external data}

\subsubsection{Uniform use of the same control treatment in the population underlying the external data}

Consider a setting where the index trial has evaluated an experimental treatment $A=1$ and a control treatment $A=0$. Suppose treatment $A=0$ is also \textit{uniformly adopted} in the population underlying the external data; that is, all individuals receive treatment $A=0$, which might reflect scenarios where there is a single standard of care for the condition under study. In this setting, depicted in Figure \ref{fig:data_structure_1}, we may be interested in using external data to improve statistical efficiency in the index trial, sometimes referred to as ``trial augmentation'' or ``borrowing strength''  from the external data  \cite{karlsson2024robust, bartolomeis2025}. We note that this is not necessary to identify the causal estimand of interest, $\E[Y^{a=1}-Y^{a=0}| S = 1]$, which is possible using index trial data alone.

\begin{figure}[H]
    \caption{Schematic representation of data structure when there is uniform adoption of a control treatment $A=0$ in the population underlying the external data.}
    \centering
\renewcommand{\arraystretch}{1.3}
\begin{tabular}{|l|l|l|l|}
\hline
\multirow{3}{*}{\quad\quad$X$\quad\quad} & \multirow{2}{*}{\;\;\;$S=1$\;\;\;} & \;\;\;$A=1$\;\;\; & \multirow{3}{*}{\quad\quad$Y$\quad\quad} \\ \cline{3-3}
 &  & \;\;\;$A=0$\;\;\; &  \\ \cline{2-3}
 & \;\;\;$S=0$\;\;\; & \;\;\;$A=0$\;\;\; &  \\ \hline
\end{tabular}
    \label{fig:data_structure_1}
\end{figure}

\paragraph{Identifiability conditions}\mbox{}

\noindent For this data structure, we require an additional identifiability condition.

\noindent 
(A6) \textit {Uniform use of treatment $A=0$ in the external data.} The probability of being assigned treatment $A = 0$ in the external data is 1. That is, for every covariate pattern $x$ with positive density in the population underlying the external data $f(x, S = 0) > 0$, $\Pr[A = 0|X = x, S = 0] = \Pr[A = 0|S = 0] = 1$. This condition implies that if $S = 0$, then $A = 0$, and thus, for $f(x, S = 0) > 0$, $\E [Y | X = x, S = 0] = \E [Y | X = x, S = 0, A = 0]$. Furthermore, this condition implies that treatment-outcome confounding is not a concern in the external data, since treatment is restricted to one level, $A=0$.

\paragraph{Identification}\mbox{}

\noindent Under this setup, we first show that there is a ``testable implication'' \cite{robins1986, robinsandwasserman1997} of the identifiability conditions, in the sense that we can examine whether the implied equality of conditional expectations for treatment $A=0$, in the common support of the densities of covariates $X$ between the populations under combination, is compatible with the observed data. This restriction on the law of the data is later used to identify the potential outcome mean under the intervention set treatment $A$ to $a=0$ in the population underlying the trial. 

\begin{prop}\label{prop2}
Conditions A1, A2, A3 and A4 for $a=0$, and condition A6, impose the following restriction on the law of the data. For every covariate pattern $x$ in the common support between the populations underlying the two data sources, that is, for every $x$ with $f(x, S=1) > 0$ and $f(x, S=0) > 0$, $$\E [Y |X = x , S = 1, A = 0] =  \E [Y | X = x,S = 0, A = 0].$$
\end{prop}

\noindent There is no requirement for any specific covariate pattern $x$ to be represented within the external data to learn about the index trial. That is, the index trial suffices to learn about potential outcome means conditional on the covariates $x$ that have a positive density in the population underlying the index trial, even if some of these covariate patterns cannot occur in the population underlying the external data. For any specific covariate pattern $x$ that has positive support in the index trial, that is $f(x,S=1)>0$, but not in the external data, $f(x,S=0)=0$, we can write $$\E [Y^{a=0}|X=x, A=0]=\E [Y^{a=0}|X=x, S=1, A=0],$$ whereas the corresponding conditional expectation in the external data need not be defined.

Next, we address the identification of the potential outcome mean under the intervention to set treatment $A$ to $a=0$ in the population underlying the index trial.

\begin{prop}\label{prop3}
Under conditions A1, A2, A3, A4 for $a=0$, and condition A6, the potential outcome mean under the intervention to set treatment $A$ to $a=0$ in the population underlying the index trial,  $\E[Y^{a=0} | S = 1]$, is identified with $$\beta \equiv \E \big[ \E [Y|X, A = 0] \big| S = 1 \big].$$ 
\end{prop}

This result combines data from individuals given the control treatment $A=0$ in the populations underlying the trial and the external data, marginalized over the covariate distribution in the trial. Using the above results, it is easy to see that the average treatment effect comparing intervention to set treatment $A$ to $a=1$ versus $a=0$ in the population underlying the trial, $\E[ Y^{a=1} - Y^{a=0} | S = 1]$, is identified using Propositions 1 and 3. 

\begin{prop}\label{prop4}
If the conditions required for Propositions 1 and 3 hold, the average treatment effect comparing the intervention to set treatment $A$ to $a=1$ versus $a=0$ in the population underlying the index trial, $\E[ Y^{a=1} - Y^{a=0} | S = 1]$, is identified by $\kappa \equiv \gamma_{1,1} - \beta$.
\end{prop}

\subsubsection{Uniform use of a third treatment in the population underlying the external data}

Now consider a setting where $A=2$ is another experimental treatment that is uniformly adopted in the population underlying the external data, but that is not evaluated in the index trial. In this setting, depicted in Figure \ref{fig:data_structure_2}, we may wish to estimate the average treatment effect comparing the experimental treatment in the trial $A=1$ and some other treatment $A=2$ in the external data, as captured by the causal estimand $\E[Y^{a=1} - Y^{a=2} | S = 1 ]$. The treatment $A=2$ is sometimes referred to as a type of ``external comparator''.

\begin{figure}[H]
    \caption{Schematic representation of data structure when there is uniform adoption of a treatment, $A=2$ in the population underlying the external data, with the treatment $A=2$ not studied in the index trial.}
    \centering
\renewcommand{\arraystretch}{1.3}
\begin{tabular}{|l|l|l|l|}
\hline
\multirow{3}{*}{\quad\quad$X$\quad\quad} & \multirow{2}{*}{\;\;\;$S=1$\;\;\;} & \;\;\;$A=1$\;\;\; & \multirow{3}{*}{\quad\quad$Y$\quad\quad} \\ \cline{3-3}
 &  & \;\;\;$A=0$\;\;\; &  \\ \cline{2-3}
 & \;\;\;$S=0$\;\;\; & \;\;\;$A=2$\;\;\; &  \\ \hline
\end{tabular}
    \label{fig:data_structure_2}
\end{figure}

\paragraph{Identifiability conditions} \mbox{}

\noindent (A6') \mbox{}
\textit {Uniform use of experimental treatment $A=2$ in the external data.} The probability of being assigned treatment $A = 2$ in the external data is 1. That is, for every covariate pattern $x$ with positive density in the population underlying the external data $f(x, S = 0) > 0$, $\Pr[A = 2|X = x, S = 0] = \Pr[A = 2|S = 0] = 1$. This condition implies that if $S = 0$, then $A = 2$, and thus for $f(x, S = 0) > 0$, $\E [Y | X = x, S = 0] = \E [Y | X = x, S = 0, A = 2]$. This condition implies that treatment-outcome confounding is not a concern in the population underlying the external data.

\paragraph{Identification}
\begin{prop}\label{prop5}
Under conditions A1 and A4 for $a=2$, and conditions A5 and A6', the potential outcome mean under the intervention to set treatment $A$ to $a=2$ in the population underlying the index trial, $\E[ Y^{a=2} | S = 1]$, is identified with $$\eta \equiv \E \big[ \E [Y|X, S = 0] \big| S = 1 \big].$$
\end{prop}

This identification result is the conditional outcome mean among individuals underlying the external data, marginalized to the covariate distribution in the population underlying the index trial. Proposition 5 requires the positivity conditions A5 and A6' because data on the treatment $A=2$ are obtained exclusively from the external data. By contrast, in Proposition 2, data on the treatment $A=0$ were sourced from the index trial and external data; as such there was no requirement for positivity condition A5.

Using the above results, the average treatment effect comparing the intervention to set treatment $A$ to $a=1$ versus $a=2$ in the population underlying the trial, $\E[ Y^{a=1} - Y^{a=2} | S = 1]$, is identified using Propositions 1 and 5.

\begin{prop}\label{prop6} If the conditions required for Propositions 1 and 5 hold, the average treatment effect comparing the intervention to set treatment $A$ to $a=1$ versus $a=2$ in the population underlying the index trial, $\E[ Y^{a=1} - Y^{a=2} | S = 1]$, is identified with $\zeta \equiv \gamma_{1,1}-\eta$.
\end{prop}

\noindent Proposition 6 can be easily modified to represent the average treatment effect if one were to combine an index single-group trial with treatment $A=1$ with external data where there is uniform adoption of a treatment $A=2$. The only change required would be the approach to identify the potential outcome mean under the intervention to set treatment $A$ to $a=1$ within the population underlying the trial. In Proposition 1, one would not require conditional exchangeability over treatment $A$ in the index trial (A2), because the treatment $A=1$ is uniformly adopted in the index single-group trial (i.e., confounding for treatment has been accounted for by restriction). Furthermore, the condition of positivity for treatment (A3) would be modified such that the probability of treatment assignment $A=1$ is 1.

\subsection{Identification under treatment variation in the population underlying the external data}

We now consider identification when there is treatment variation in the population underlying the external data. Suppose the population underlying the external data comprises individuals who may be assigned treatments, including $A=2$ and some other treatment, as depicted in Figures \ref{fig:data_structure_3} and \ref{fig:data_structure_4}. The treatment $A=2$ is also sometimes described as an ``external comparator'' to the index trial, which has only evaluated $A=1$ and $A=0$. Here, one may consider two identification approaches to estimate the average treatment effect comparing the intervention to set treatment $A$ to $a=1$ versus $a=2$ in the population underlying the index trial, $\E[Y^{a=1} - Y^{a=2} | S = 1 ]$. With respect to trial participation $S$, the first is based on conditional exchangeability (transportability) in conditional outcome means, and the second uses conditional exchangeability (transportability) in effect measures.

\begin{figure}[H]
    \caption{Schematic representation of data structure when there is treatment variation in the population underlying the external data, within which there is $A=2$ and some other treatment $A$ that was not studied in either the index or external data.}
    \centering
\renewcommand{\arraystretch}{1.3}
\begin{tabular}{|l|l|c|l|}
\hline
\multirow{4}{*}{\quad\;\;\;$X$\;\;\;\quad} & \multirow{2}{*}{\;\;\;$S=1$\;\;\;} & \;\;\;$A=1$\;\;\; & \multirow{4}{*}{\quad\;\;\;$Y$\;\;\;\quad} \\ \cline{3-3}
 &  & \;\;\;$A=0$\;\;\; &  \\ \cline{2-3}
 & \multirow{2}{*}{\;\;\;$S=0$\;\;\;} & $A=2$ &  \\ \cline{3-3}
 &  & $A \notin \{0,1,2\}$ &  \\ \hline
\end{tabular}
    \label{fig:data_structure_3}
\end{figure}

\begin{figure}[H]
    \caption{Schematic representation of data structure when there is treatment variation in the population underlying the external data, within which there is $A=2$ and $A=0$. Here, $A=0$ is common to both $S=1$ and $S=0$.}
    \centering
\renewcommand{\arraystretch}{1.3}
\begin{tabular}{|l|l|c|l|}
\hline
\multirow{4}{*}{\quad\quad$X$\quad\quad} & \multirow{2}{*}{\;\;\;$S=1$\;\;\;} & \;\;\;$A=1$\;\;\; & \multirow{4}{*}{\quad\quad$Y$\quad\quad} \\ \cline{3-3}
 &  & \;\;\;$A=0$\;\;\; &  \\ \cline{2-3}
 & \multirow{2}{*}{\;\;\;$S=0$\;\;\;} & $A=2$ &  \\ \cline{3-3}
 &  & $A = 0$ &  \\ \hline
\end{tabular}
    \label{fig:data_structure_4}
\end{figure}

\subsubsection{Identification under transportability in mean}

First, we consider identification based on transportability in mean. The setting here does not require a common treatment between the data sources, in the sense that the second treatment in the external data need not necessarily be $A=0$. Relevant data structures are shown in Figures \ref{fig:data_structure_3} and \ref{fig:data_structure_4}.

\paragraph{Identifiability conditions} \mbox{}

\noindent (A7) \textit {Conditional exchangeability over treatment $A$ in the population underlying the external data.} For the intervention to set treatment $A$ to $a\in \mathcal{A}_{S=0}$, $\E[Y^{a} | X = x , S = 0 ] = \E[Y^{a} | X = x, S = 0 , A = a]$. This condition is implied by the independence conditions $Y^{a} \indep A | (X , S = 0)$ and $(Y^{a}, X) \indep A | S = 0$, which would be supported by study design if the external data are obtained from a marginally randomized trial.\\

\noindent (A8) \textit {Positivity of treatment in the population underlying the external data.} For every $a \in \mathcal A_{S = 0}$, and every covariate pattern $x$ with positive density in the external data, the probability of treatment assignment is positive. That is, if $f(x, S = 0) > 0$, then $\Pr[ A = a | X = x , S = 0 ]>0$. Stated differently, in the external data there is a non-zero probability of being assigned to the treatment, conditional on the covariates $X$ required for conditional exchangeability over treatment $A$.   

\paragraph{Identification}
\begin{prop} \label{prop7}
Under conditions A1, A4, A7, and A8 for $a=2$, and condition A5, the potential outcome mean under the intervention to set treatment $A$ to $a=2$ in the population underlying the index trial, $\E[ Y^{a=2}| S = 1]$, is identified with $$\gamma_{0,2} \equiv \E \big[ \E [Y|X, S = 0, A = 2] \big| S = 1 \big].$$
\end{prop} 

To obtain the identification result $\gamma_{0,2}$, there is no requirement to use the second treatment in either the index trial ($S=1$) or external data ($S=0$). Therefore, it is not necessary for any common treatments to be adopted in both populations underlying the data sources. Furthermore, identification requires an assumption about the absence of confounding for treatment in the population underlying the external data, as was the case in Proposition 5. Though the absence of confounding was expected in Proposition 5 because treatment only had one level, in Proposition 7 one must invoke conditional exchangeability for treatment in the external data (A7).

Using the above results, the average treatment effect comparing intervention to set treatment $A$ to $a=1$ versus $a=2$ in the population underlying the trial, $\E[ Y^{a=1} - Y^{a=2} | S = 1]$, is identified using Propositions 1 and 7.

\begin{prop} \label{prop8}
If the conditions required for Propositions 1 and 7 hold, the average treatment effect comparing the intervention to set treatment $A$ to $a=1$ versus $a=2$ in the population underlying the index trial, $\E[ Y^{a=1} - Y^{a=2} | S = 1]$, is identified with $ \psi \equiv \gamma_{1,1}-\gamma_{0,2}.$
\end{prop}

Proposition 8 is easily modified to represent the average treatment effect if one were to combine an index single-group trial with treatment $A=1$ with external data where there is treatment variation (including $A=2$ and some other treatment). However, identification would not require conditional exchangeability for treatment (A2), and the probability of treatment assignment would be 1, therefore requiring a small amendment to A3. 

\subsubsection {Identification under transportability in effect measure}

We now consider a treatment comparison between treatment $A=1$ in the index trial and treatment $A=2$ in the external data, using treatment $A=0$ as a shared comparator in both data sources. Such settings, depicted in Figure \ref{fig:data_structure_4}, may allow for treatment comparisons under transportability in mean. However, by anchoring the comparison on the shared treatment $A=0$, identification is also possible by invoking conditional exchangeability (transportability) in difference or relative effect measures over trial participation $S$ rather than in means (A4). The condition of transportability in effect measure is weaker than transportability in mean, because the latter implies the former, but the converse is not necessarily true. However, the assumption presupposes that the common treatment $A=0$ is truly common between the index and external data. Nontrivial differences (e.g., with respect to differential blinding) may violate consistency because version irrelevance does not hold.

\paragraph{Transportability of difference measures} \mbox{}

\noindent First, suppose the investigator has substantive knowledge to support the assumption that difference effect measures are transportable between the populations underlying the index trial ($S=1$) and external data ($S=0$), conditional on covariates $X$. We show that under this assumption, the causal estimand of interest $\E[Y^{a=1} - Y^{a=2} | S = 1 ]$ can be identified in a manner where the conditional outcome mean for the common treatment $A=0$ is not required to be equal in the populations underlying the trial and the external data.

\subparagraph*{Identifiability conditions}\mbox{}

\noindent (A9) \textit {Conditional exchangeability (transportability) of difference effect measures over trial participation $S$.} For the intervention setting treatment $A$ to $a$ and $a'$, and for every covariate pattern $x$ with positive densities $f(x,S=1) > 0 $ and $f(x,S=0) > 0 $, $\E[Y^{a} - Y^{a'} | X = x , S = 1] = \E[Y^{a} - Y^{a'} | X = x, S = 0]$. This condition implies no effect modification by $S$ for the effect of the intervention to set treatment $A$ to $a$ versus $a'$ on the outcome on the difference scale, conditional on covariates $X$.

\subparagraph*{Identification}

\begin{prop}\label{prop9}
Under conditions A1, A2, A3, and A7 for $a=0$; conditions A1 and A7 for $a=2$; and conditions A5, A8 for $a=0,2$, and A9 for $a=0,2$, the potential outcome mean under the intervention to set treatment $A$ to $a=2$ in the population underlying the index trial, $\E[ Y^{a=2} | S = 1]$, is identified with $\lambda \equiv \gamma_{0,2}+\left(\gamma_{1,0}-\gamma_{0,0}\right)$.
\end{prop}

\noindent In $\lambda$, the expression $\gamma_{1,0}-\gamma_{0,0}$ can be thought of as a correction of $\gamma_{0,2}$. This correction uses information about differences between the populations underlying the index and external data, specifically the conditional outcome means under the common control treatment, averaged over the covariate distribution of the population underlying the index trial. 

Using the above results, the average treatment effect comparing intervention to set treatment $A$ to $a=1$ versus $a=2$ in the population underlying the trial, $\E[ Y^{a=1} - Y^{a=2} | S = 1]$, is identified using Propositions 1 and 9.

\begin{prop}\label{prop10}
If the conditions required for Propositions 1 and 9 hold, the average treatment effect comparing intervention to set treatment $A$ to $a=1$ versus $a=2$ in the population underlying the index trial, $\E[ Y^{a=1} - Y^{a=2} | S = 1]$, is identified with $\phi\equiv\gamma_{1,1}-\lambda$. That is, $$\phi=\gamma_{1,1}-\gamma_{1,0}-\left(\gamma_{0,2}-\gamma_{0,0}\right).$$
\end{prop}

Proposition 10 represents the difference in the average treatment effect comparing the populations underlying the trial and the external data, a result parallel to some ``difference-in-differences'' identification strategies in other contexts \cite{athey2006identification, sofer2016negative}. Here, the identifying expression for the average treatment effect represents the difference of two conditional outcome means marginalized to the covariate distribution in the population underlying the index trial.

\paragraph{Transportability of relative measures}\mbox{}

\noindent Now suppose the investigator has substantive knowledge to support the assumption that relative effect measures are transportable between the populations underlying the trial and the external data, conditional on covariates $X$ \cite{schwartz2006ratio, spiegelman2017evaluating, huitfeldt2019collapsibility, huitfeldt2019effect, dahabreh2024relative}. 

\subparagraph*{Identifiability conditions} \mbox{}

\noindent
(A10) \textit{Conditional exchangeability (transportability) of relative effect measures over trial participation $S$.} For every covariate pattern $x$ with positive densities $f(x,S=1)>0$, and for the intervention setting treatment $A$ to $a$ and $a'$, $$ \dfrac{\E[Y^{a} | X = x , S = 1]}{ \E [ Y^{a'} | X = x, S = 1]} = \dfrac{\E[Y^{a}  | X = x , S = 0] }{ \E[Y^{a'}  | X = x , S = 0]},$$ with $\E [ Y^{a} | X = x, S = s] \neq 0$ and $\E[Y^{a'}  | X = x , S = s] \neq 0$ for $s=0,1$.

\subparagraph*{Identification}

\begin{prop}\label{prop11}
Under conditions A1, A2, and A3 for $a=0$, conditions A1 and A7 for $a=2$, and conditions A5, A8 for $a=0,2$, and A10 for $a=0,2$, the potential outcome mean under intervention to set treatment $A$ to $a=2$ in the population underlying the index trial, $\E[Y^{a=2}|S=1]$, is identified with $$\rho \equiv \E \left [\E[Y|X, S=1, A=0] \dfrac{\E[Y| X , S = 0, A=2]}{\E[Y|X , S = 0, A=0]} \biggr|S=1 \right ].$$
\end{prop} 

Here, the ratio of conditional means \small$\dfrac{\E [Y | X, S = 0,  A = 2 ]}{\E[Y  | X, S = 0, A = 0]}$ \normalsize from the population underlying the external data is multiplied by $\E[Y  | X , S = 1, A = 0] $, which can be thought of as a conditional outcome mean under the control treatment $A=0$ in the trial, prior to being marginalized over the covariate distribution in the population underlying the index trial. 

Using the above results, the average treatment effect comparing intervention to set treatment $A$ to $a=1$ versus $a=2$ in the population underlying the trial, $\E[ Y^{a=1} - Y^{a=2} | S = 1]$, is identified using Propositions 1 and 11.

\begin{prop}\label{prop12} If the conditions required for Propositions 1 and 11 hold, the average treatment effect comparing intervention to set treatment $A$ to $a=1$ versus $a=2$ in the population underlying the index trial, $\E[ Y^{a=1} - Y^{a=2} | S = 1]$, is identified with $\theta \equiv \gamma_{1,1}-\rho.$
\end{prop}

In general, transportability for both absolute and relative measures is not likely to hold simultaneously, at least in the absence of extreme assumptions \cite{dahabreh2024relative, ung2025generalizing}. Thus, investigators will need to use their subject matter expertise to determine the scale, if any, on which effect measures may be transported between populations. Furthermore, we note that the structure of identification strategies that rely on transportability in effect measure, unlike that when invoking transportability in mean, permits the relaxation of the strong assumption of absent trial engagement effects (see Section 5.5). 

\subsection{Using additional covariates to control confounding in the population underlying the external data}

Until now, we have assumed that the same covariates $X$ are sufficient to achieve conditional exchangeability over treatment $A$. However, investigators may believe that conditional exchangeability with respect to treatment $A$ is only plausible when conditioning on a wider set of covariates. Suppose investigators have collected another set of baseline covariates, $W$, in the population underlying the external data, and wish to conduct a comparison under transportability in mean. Here, we consider an expanded data structure shown in Figure \ref{fig:data_structure_5}.

\begin{figure}[H]
    \caption{Schematic representation of data structure when there is treatment variation in the population underlying the external data, within which there is $A=2$ and some other treatment $A$ that was not studied in either the index or external data. Additional baseline covariates $W$ are in the external data, but not required in the index trial.}\centering
\renewcommand{\arraystretch}{1.3}
\begin{tabular}{|l|c|l|c|l|}
\hline
\multirow{4}{*}{\quad\quad$X$\quad\quad} & \multirow{2}{*}{\quad Not required \quad} & \multirow{2}{*}{\;\;\;$S=1$\;\;\;} & \;\;\;$A=1$\;\;\; & \multirow{4}{*}{\quad\quad$Y$\quad\quad} \\ \cline{4-4}
 &  &  & \;\;\;$A=0$\;\;\; &  \\ \cline{2-4}
 & \multirow{2}{*}{\quad$W$\quad} & \multirow{2}{*}{\;\;\;$S=0$\;\;\;} & $A=2$ &  \\ \cline{4-4}
 &  &  & \;\;\;$A \notin \{0,1,2\}$\;\;\; &  \\ \hline
\end{tabular}
    \label{fig:data_structure_5}
\end{figure}

\subsubsection{Identifiability conditions}

\noindent 
(A7') \textit {Conditional exchangeability over treatment $A$ in the population underlying the external data.} For every covariate pattern $(x,w)$ with positive densities in the external population $f(x,w,S=0) > 0$ and for the intervention to set treatment $A$ to $a$, $\E[Y^{a} |X = x , W = w, S = 0 ] = \E[Y^{a} | X = x, W = w, S = 0, A = a]$. This partial exchangeability condition is implied by the independence conditions $Y^{a} \indep A | (X, W, S = 0)$ and $(Y^{a}, X, W) \indep A | S = 0$, which is supported by design if the external data are from a marginally randomized trial.\\

\noindent 
(A8') \textit {Positivity of treatment in the population underlying the external data.} For treatment $a \in \mathcal A_{ S = 0 }$, and every covariate pattern $(x,w)$ with positive density in the external data, the probability of treatment assignment is positive. That is, if $f(x, w, S = 0) > 0$, then $\Pr[ A = a | X = x, W = w, S = 0 ]>0$. Stated differently, in the external data, there is a non-zero probability of being assigned to the treatment, conditional on the covariates $X$ and $W$ required for conditional exchangeability over treatment $A$. 

\subsubsection{Identification}

\begin{prop}\label{prop13}
Under conditions A1, A4, A7', and A8' for $a=2$, and A5, the potential outcome mean under the intervention to set treatment $A$ to $a=2$ in the population underlying the index trial, $\E[ Y^{a=2}| S = 1]$, is identified with $$\mu \equiv \E \left[ \E \big [ \E [Y|X, W, S = 0, A = 2] \big| X, S=0 \big] \Big| S = 1 \right].$$ 
\end{prop} 

When additional confounding adjustment is required for covariates $W$ in the population underlying the external data, the average treatment effect comparing intervention to set treatment $A$ to $a=1$ versus $a=2$ in the population underlying the index trial, $\E[ Y^{a=1} - Y^{a=2} | S = 1]$, is identified using Propositions 1 and 13.

\begin{prop}\label{prop14}
If the conditions required for Propositions 1 and 13 hold, the average treatment effect comparing the intervention to set treatment $A$ to $a=1$ versus $a=2$ in the population underlying the index trial, $\E[ Y^{a=1} - Y^{a=2} | S = 1]$, is identified with $\xi \equiv \gamma_{1,1}-\mu$.
\end{prop}

Here, the identification result $\xi$ does not require data on $W$ in the population underlying index trial, $S=1$. It suffices to have data on $W$ to control confounding for the effect of $A$ on $Y$ in the population underlying the external data.

\subsection{Identification in the presence of study engagement effects}

In applied settings, the condition of absent study engagement effects is often not plausible. For example, if participating in the index trial involves ancillary treatments or procedures that are not provided in the population underlying the external data, one may not be able to disentangle the effect of the treatment from trial-associated interventions \cite{dahabreh2019identification}. In this section, we propose an approach that can salvage identification when such effects are present, relying on the key assumption of \textit{no casual interaction} \cite{vanderweele2009b, hernan2024} between trial participation $S$ and treatment $A$ \cite{ung2025generalizing}. Interestingly, the plausibility of this assumption could in principle be assessed empirically, for example using factorial trials that randomly assign two or more treatments. 

\subsubsection{Causal estimands} We are interested in potential outcome means under joint intervention to set trial participation $S$ to $s=1$ and treatment $A$ to $a \in \mathcal{A}_{S=1} \cup \mathcal{A}_{S=0}$, that is $\E[Y^{s=1,a}|S=1]$. We are also interested in their causal mean difference, $\E[Y^{s=1,a=1}-Y^{s=1,a=2}|S=1]$. For these estimands, the interventions to set trial participation $S$ to $s=1$ and $s=0$ represent the scale up of non-treatment activities within the populations underlying the index and external data, respectively. Recall that while all prior estimands also explicitly intervened on trial participation $S$, the effect of any such intervention was assumed to be mediated entirely by treatment $A$, therefore allowing for the counterfactual $s$ superscript to be omitted from their causal estimands.

\subsubsection{Identifiability conditions} 
First, we provide identifiability conditions that are similar to those presented previously, except the effect of intervening to set trial participation $S$ to $s=0,1$ is now retained in the counterfactual superscript for the outcome $Y$.\\

\noindent (A1$^\dagger$) \textit {Consistency of potential outcomes with respect to intervention on trial participation status $S$ and treatment $A$}. If $S_i = s$ and $A_i = a$, then $Y_i^{s,a} = Y_i$. That is, the potential outcome for the $i^{th}$ individual under intervention to set trial participation $S$ to $s$ and treatment $A$ to $a$ is the observed outcome for that individual under the same observed trial participation status and treatment level.\\

\noindent (A2$^\dagger$) \textit {Conditional mean exchangeability over treatment $A$ for joint intervention to set trial participation $S$ to $s$ and treatment $A$ to $a$ within the population underlying the index trial.} For intervention to set trial participation $S$ to $s=1$, treatment $A$ to $a \in \mathcal A_{S=1}$, and every $x$ with positive densities $f(x,S=1) >0$, $\E[Y^{s=1,a} | X , S = 1] = \E[Y^{s=1,a}|X, S = 1, A=a].$\\

\noindent (A7$^\dagger$) \textit {Conditional mean exchangeability over treatment $A$ for joint intervention to set trial participation $S$ to $s$ and treatment $A$ to $a$ in the population underlying the external data.} For intervention to set trial participation $S$ to $s=0$ and treatment $A$ to $a \in \mathcal A_{S=0}$, and every $x$ with positive densities $f(x,S=0) >0$, $\E[Y^{s=0,a} | X , S = 0] = \E[Y^{s=0,a}|X, S = 0, A=a].$\\

\noindent (A9$^\dagger$) \textit {Conditional exchangeability of difference effect measures over trial participation $S$.} For intervention to set trial participation $S$ to $s=0$, treatment $A$ to $a,a'$, and every $x$ with positive densities $f(x,S=1) >0$ and $f(x,S=0)>0$, $\E[Y^{s=0,a} - Y^{s=0,a'} | X , S = 1] = \E[Y^{s=0, a} - Y^{s=0, a'} | X , S = 0].$ \\

\noindent (A10$^\dagger$) \textit {Conditional exchangeability of relative effect measures over trial participation $S$.}  For intervention to set trial participation $S$ to $s=0$, treatment $A$ to $a,a'$, and every $x$ with positive densities $f(x,S=1) >0$ and $f(x,S=0)>0$, $$\dfrac{\E[Y^{s=0,a} | X , S = 1]}{\E[Y^{s=0,a'} | X , S = 1]}=\dfrac{\E[Y^{s=0,a} | X , S = 0]}{\E[Y^{s=0,a'} | X , S = 0]}.$$

\noindent (A11) \textit {No causal interaction between trial participation $S$ and treatment $A$ on the causal risk difference scale.}  
For treatments $a,a'$ and every $x$ with positive densities $f(x, S=1) >0$ and $f(x, S=0)>0$, $$\E[Y^{s=1,a} - Y^{s=1,a'} | X , S = 1] = \E[Y^{s=0, a} - Y^{s=0, a'} | X , S = 1].$$ That is, within the population underlying the index trial, there is no causal interaction \cite{vanderweele2014interaction, hernan2024} between trial participation $S$ and treatment $A$ on the causal risk difference scale, conditional on covariates $X$.\\

\noindent (A11') \textit {No causal interaction between trial participation $S$ and treatment $A$ on the causal risk ratio scale.}  
For treatments $a,a'$, trial participation status $s = 0,1,$ and every $x$ with positive densities $f(x, S=1) >0$ and $f(x, S=0)>0$, $$\dfrac{\E[Y^{s=1,a} | X , S = 1] }{\E[Y^{s=1,a'}| X , S = 1]}= \dfrac{\E[Y^{s=0,a}| X , S = 1] }{\E[Y^{s=0,a'} | X , S = 1]}.$$

\subsubsection{Identification} 

First, we consider identification of the potential outcome mean using conditions only in the index trial.

\begin{prop}\label{prop15}
Under conditions A1$^\dagger$ for $s=1,a=1$; and A2$^\dagger$ and A3 for $a=1$, the potential outcome mean under intervention to set trial participation $S$ to $s=1$ and treatment $A$ to $a=1$ in the population underlying the index trial, $\E[ Y^{s=1,a=1}| S = 1]$, is identified with $\gamma_{1,1}$.
\end{prop}

Next, we consider identification of the potential outcome mean under intervention to set trial participation $S$ to $s=1$ and treatment $A$ to $a=2$, in the population underlying the index trial. First, we consider identification under a no causal interaction assumption on the difference scale.

\begin{prop}\label{prop16}
Suppose the following conditions hold: A1$^\dagger$ for $(s=0,a=0)$, $(s=0,a=2)$, and $(s=1,a=0)$; A2$^\dagger$ and A3 for $a=0$; A5; and A7$^\dagger$, A8, A9$^\dagger$, and A11 for $a=0,2$. The potential outcome mean under intervention to set trial participation $S$ to $s=1$ and treatment $A$ to $a=2$ in the population underlying the index trial, $\E[ Y^{s=1, a=2} | S = 1]$, is identified with $\lambda \equiv \gamma_{0,2}+\left(\gamma_{1,0}-\gamma_{0,0}\right).$
\end{prop}

Using this result, the average treatment effect comparing intervention to set trial participation $S$ to $s=1$ and treatment $A$ to $a=1$ versus trial participation $S$ set to $s=1$ and treatment $A$ to $a=2$ within the population underlying the index trial follows immediately.

\begin{prop}\label{prop17}
If the conditions required for Propositions 15 and 16 hold, the average treatment effect in the population underlying the trial, comparing joint intervention to set treatment $A$ to $a=1$ versus $a=2$ had trial participation $S$ also been set to $s=1$, $\E[ Y^{a=1} - Y^{a=2} | S = 1]$, is identified with $\phi\equiv \gamma_{1,1}-\gamma_{1,0}-\left(\gamma_{0,2}-\gamma_{0,0}\right).$
\end{prop}

We now consider identification under no causal interaction on the ratio scale.

\begin{prop}\label{prop18}
Suppose the following conditions hold: A1$^\dagger$ for $(s=0,a=0)$, $(s=0,a=2)$, and $(s=1,a=0)$; A2$^\dagger$ and A3 for $a=0$; A5; and A7$^\dagger$, A8, A10 $^\dagger$, and A11 for $a=0,2$. The potential outcome mean under intervention to set trial participation $S$ to $s=1$ and treatment $A$ to $a=2$ in the population underlying the index trial, $\E[ Y^{s=1, a=2} | S = 1]$, is identified with $\rho$.
\end{prop}

Using this result, the average treatment follows immediately.

\begin{prop}\label{prop19}
If the conditions required for Propositions 15 and 18 hold, the average treatment effect in the population underlying the trial, comparing joint intervention to set treatment $A$ to $a=1$ had trial participation $S$ been set to $s=1$, $\E[ Y^{a=1} - Y^{a=2} | S = 1]$, is identified with $\theta\equiv\gamma_{1,1}-\rho$.
\end{prop}

The identification results here are the same as the observed data functionals obtained under transportability in effect measure, as seen in Propositions 1 and 9 through 12. This suggests that the estimators constructed using the observed data functional $\phi$ and $\theta$ may have different interpretations, depending on the extent to which trial engagement effects are believed to be present for the causal problem at hand.

\subsection{Identification when the index trial and external data are used to learn about other target populations}

Suppose that investigators wish to combine information from an index trial with external data to learn about causal effects in another target population captured in a third data source. This target population is also external to the index trial, but we clarify our slight abuse in terminology here to avoid confusion. Here, the external comparator analysis may be viewed as being nested within a classic generalizability or transportability problem, and may be of particular interest when the third dataset represents a target population in which the treatments would be expected to be deployed in real-life settings \cite{dahabreh2020toward, dahabreh2023efficient}. 

The data structures considered here are depicted in Figure \ref{fig:data_structure_6} and Figure \ref{fig:data_structure_7}. The data structure in Figure \ref{fig:data_structure_6} involves only baseline covariates within the $S=2$ population, with no need for information on treatments $A$ nor outcomes $Y$. The data structure in Figure \ref{fig:data_structure_7} contains information on baseline covariates $X$, the control treatment $A=0$, and outcome $Y$, and is reserved when transporting relative measures from the index and external data to the $S=2$ population.

\begin{figure}[H]
    \caption{Schematic representation of data structure including the index trial, external data, and a third population $S=2$ which contains only baseline covariate information.}
    \centering
\renewcommand{\arraystretch}{1.3}
\begin{tabular}{|l|l|c|l|}
\hline
\multirow{5}{*}{\quad\quad$X$\quad\quad} & \multirow{2}{*}{\;\;\;$S=1$\;\;\;} & \;\;\;$A=1$\;\;\; & \multirow{4}{*}{\quad\quad\quad\quad$Y$\quad\quad} \\ \cline{3-3}
 &  & \;\;\;$A=0$\;\;\; &  \\ \cline{2-3}
 & \multirow{2}{*}{\;\;\;$S=0$\;\;\;} & $A=2$ &  \\ \cline{3-3}
 &  & $A = 0$ &  \\ \cline{2-4}
 & \;\;\;$S=2$\;\;\; & Not required & \quad\quad Not required \quad\quad \\ \hline
\end{tabular}
    \label{fig:data_structure_6}
\end{figure}

\begin{figure}[H]
    \caption{Schematic representation of data structure including the index trial, external data, and a third population $S=2$ which contains covariate, treatment, and outcomes under the control treatment $A=0$.}
    \centering
\renewcommand{\arraystretch}{1.3}
\begin{tabular}{|l|l|c|l|}
\hline
\multirow{5}{*}{\quad\quad$X$\quad\quad} & \multirow{2}{*}{\;\;\;$S=1$\;\;\;} & \;\;\;$A=1$\;\;\; & \multirow{5}{*}{\quad\quad$Y$\quad\quad} \\ \cline{3-3}
 &  & \;\;\;$A=0$\;\;\; &  \\ \cline{2-3}
 & \multirow{2}{*}{\;\;\;$S=0$\;\;\;} & $A=2$ &  \\ \cline{3-3}
 &  & $A = 0$ &  \\ \cline{2-3}
 & \;\;\;$S=2$\;\;\; & $A=0$ &  \\ \hline
\end{tabular}
    \label{fig:data_structure_7}
\end{figure}

\subsubsection{Causal estimands} Under non-nested sampling, where all data sources are collected separately, we are interested in potential outcome means under intervention to set treatment $A$ to $a=1$ and $a=2$, within a target population underlying some third data source, represented by $S=2$, that is $\E[Y^{a}|S=2]$ for $a=1,2$.  We are also interested in their causal contrast on the difference scale, represented by $\E[Y^{a=1}-Y^{a=2}|S=2]$.

\subsubsection{Identifiability conditions} The conditions used here are similar to those already presented, with abridged versions provided below.\\

\noindent (A4$^*$) \textit {Conditional mean exchangeability (transportability) over trial participation $S$.} For covariate patterns $f(x, S=2) > 0$, trial participation $S$ levels $s=0,1$, and treatments $a \in \mathcal A_{S=1} \cup \mathcal A_{S=0}$, $\E[Y^a|X = x, S = 2] = \E[Y^a|X = x, S = s]$.\\

\noindent (A5$^*$) \textit{Positivity for the population underlying the third data source.} That is, if $f(x, S = 2) > 0$, then $\Pr[ S = s | X = x ] > 0$ for $s=0,1$. \\

\noindent (A7$^*$) \textit {Conditional exchangeability over treatment $A$ in the population underlying the third data source.} For the intervention to set treatment $A$ to $a\in\mathcal{A}_{S=2}$, $\E[Y^{a} | X = x , S = 2 ] = \E[Y^{a} | X = x, S = 2 , A = a]$.\\

\noindent (A8$^*$) \textit {Positivity of treatment in the population underlying the third data source.} For every $a \in \mathcal A_{S = 2}$, and every covariate pattern $x$ with positive density in the third data source, the probability of treatment assignment is positive. That is, if $f(x, S = 2) > 0$, then $\Pr[ A = a | X = x , S = 2 ]>0$.\\

\noindent (A9$^*$) \textit {Conditional exchangeability (transportability) of difference effect measures over trial participation $S$.}  For covariate patterns $f(x, S=2) > 0$, trial participation $S$ levels $s=0,1$, and treatments $a \in \mathcal A_{S=1} \cup \mathcal A_{S=0}$, $\E[Y^{a} - Y^{a'} | X = x , S = 2] = \E[Y^{a} - Y^{a'} | X = x, S = s]$.\\

\noindent (A10$^*$) \textit{Conditional exchangeability (transportability) of relative effect measures over trial participation $S$.} For every covariate pattern $x$ with positive densities $f(x,S=2)>0$, and treatments $a \in \mathcal A_{S=1} \cup \mathcal A_{S=0}$, 
$$ \dfrac{\E[Y^{a} | X = x , S = 2]}{ \E [ Y^{a'} | X = x, S = 2]} = \dfrac{\E[Y^{a}  | X = x , S = s] }{ \E[Y^{a'}  | X = x , S = s]},$$ with $\E [ Y^{a} | X = x, S = s] \neq 0$ and $\E[Y^{a}  | X = x , S = s] \neq 0$ for $s=0,1$.

\subsubsection{Identification}

First, we address identification of the potential outcome mean under intervention to set treatment $A$ to $a=1$ in the population underlying the third data source, $S=2$.

\begin{prop}\label{prop20}
Under conditions A1, A2, A3 for $a=1$; A4$^*$ for $s=1$ and $a=1$; and A5$^*$ for $s=1$, the potential outcome mean under intervention to set treatment $A$ to $a=1$ in the population underlying the third data source, $\E[Y^{a=1}| S = 2]$, is identified with $$\gamma_{1,1}^* \equiv \E\left[\E[Y|X,S=1,A=1]\big|S=2\right].$$   
\end{prop}

This result is similar to the statistical estimand $\gamma_{1,1}$ presented in Proposition 1, except the conditional outcome mean under treatment $A=1$ in the index trial is marginalized to the covariate distribution in the target population underlying the third data source. 

Next, we address identification of the potential outcome mean under intervention to set treatment $A$ to $a=2$ in the population underlying the third data source.

\begin{prop}\label{prop21}
Under conditions A1, A7, and A8 for $a=2$; A4$^*$ for $a=2$ and $s=0$, and A5$^*$ for $s=0$, the potential outcome mean under intervention to set treatment $A$ to $a=2$ in the population underlying the third data source, $\E[Y^{a=2}| S = 2]$, is identified with $$\gamma_{0,2}^* \equiv \E\left[\E[Y|X,S=0,A=2]|S=2\right].$$  
\end{prop}

This result marginalizes the conditional outcome mean under treatment $A=2$ in the external data to the covariate distribution in the population underlying the third data source. Using the two results above, the causal contrast comparing intervention to set treatment $A$ to $a=1$ versus $a=2$ in the population underlying the third data source is immediate.

\begin{prop}\label{prop22} If the conditions required for Propositions 20 and 21 hold, the average treatment effect comparing intervention to set treatment $A$ to $a=1$ versus $a=2$ in the population underlying the third data source, $\E[ Y^{a=1} - Y^{a=2} | S = 2]$, is identified with $\psi^*\equiv\gamma_{1,1}^*-\gamma_{0,2}^*$.   
\end{prop}

Next, we address identification under transportability in difference effect measure, beginning with the difference scale. Here, because the goal is to not invoke any mean transportability assumptions, we will work with the causal effect comparing intervention to set treatment $A$ to $a=1$ vs. $a=2$ in the population underlying the third data source, $\E[Y^{a=1}-Y^{a=2}| S = 2]$

\begin{prop}\label{prop23} Suppose the following conditions hold: A1 for all values of $a$; A2 and A3 for $a=0,1$; A5$^*$ for $s=0,1$; A7 and A8 for $a=0,2$; A9$^*$ for $s=1$, and $a=0,1$ and $A9^*$ for $s=0$, $a=0,2$. the average treatment effect comparing intervention to set treatment $A$ to $a=1$ versus $a=2$ in the population underlying the third data source, $\E[ Y^{a=1} - Y^{a=2} | S = 2]$, is given by \vspace{-0.1in}

\noindent\resizebox{\columnwidth}{!}{%
\begin{minipage}{\columnwidth}
\begin{align*}
   \begin{split}
    \phi^* &\equiv \Big\{ \E \big[ \E [Y | X, S = 1, A = 1] \big| S = 2 \big]  - \E \big[ \E [Y | X, S = 1, A = 0 ] \big| S = 2 \big] \Big\} \\
      &\quad\quad\quad\quad\quad- \Big\{ \E \big[ \E [Y | X, S = 0, A = 2 ] \big| S = 2 \big]  - \E \big[ \E [Y | X, S = 0, A = 0] \big| S = 2 \big] \Big \} \\
    \end{split}
\end{align*}
\end{minipage}}
\end{prop}

Interestingly, identification here does not impose a testable restriction on the law of the observed data in the population underlying the third data source, specifically for the treatment $A=0$ common to the index and external data. That is, identification here does not impose the restriction that $\E\left[\E[Y|X,S=1, A=0]\big|S=2\right]=\E\left[\E[Y|X,S=0, A=0]\big|S=2\right].$

Last, we address identification under transportability in relative effect measure. Under this assumption, one requires information on the conditional outcome mean (i.e., the baseline risk) under the control treatment $A=0$ within the $S=2$ population, as seen in Table \ref{fig:data_structure_7}. First, we address identification of the potential outcome mean under intervention to set treatment $A$ to $a=1$ in the population underlying the third data source, $S=2$.

\begin{prop}\label{prop24}
Suppose the following conditions hold: A1, A2, A3 for $a=0,1$; A5$^*$ for $s=1$; A7$^*$ and A8$^*$ for $a=0$; and A10$^*$ for $s=1$ and $a=0,1$. The potential outcome mean under intervention to set treatment $A$ to $a=1$ in the population underlying the third data source, $\E[Y^{a=1} | S = 2]$, is given by $$ \rho^*_1\equiv\E \left [\E[Y|X, S=2, A=0]  \dfrac{\E[Y| X , S = 1, A=1]}{\E[Y|X, S = 1, A=0]} \biggr|S=2 \right ].$$ \end{prop}

Using symmetrical arguments, we can identify the potential outcome mean under intervention to set treatment $A$ to $a=2$ in the population underlying the third data source.

\begin{prop}\label{prop25} Suppose the following conditions hold: A1, A7, A8 for $a=0,2$; A5$^*$ for $s=0$; A7$^*$ and A8$^*$ for $a=0$; and 
A10$^*$ for $s=0$ and $a=0,2$. The potential outcome mean under intervention to set treatment $A$ to $a=2$ in the population underlying the third data source, $\E[Y^{a=2} | S = 2]$, is given by $$\rho^*_2\equiv\E \left [\E[Y|X, S=2, A=0]  \dfrac{\E[Y| X , S = 0, A=2]}{\E[Y|X, S = 0, A=0]} \biggr|S=2 \right ].$$  \end{prop}

The average causal effect obtained by transporting relative effect measures from the index and external data to the the population underlying the third data source is immediate. 

\begin{prop}\label{prop26} If the conditions required for Propositions 24 and 25 hold, the average treatment effect comparing the intervention to set treatment $A$ to $a=1$ versus $a=2$ in the third target population of interest, $\E[ Y^{a=1} - Y^{a=2} | S = 2]$, is identified with $\nu=\rho_1^*-\rho_2^*$.   
\end{prop}

\section{ESTIMATION AND INFERENCE}

There is now an extensive literature on statistical methods for the basic study templates we have described \cite{cole2010, westreich2017, dahabreh2019relation, dahabreh2019generalizingbiometrics, ventz2022, li2023improving, dahabreh2023efficient, shook2024fusing, karlsson2024robust, valancius2024causal, bartolomeis2025, li2025data}, with varying degrees of engagement with issues of sampling, the need to explicitly specify causal estimands, and identification. In general, the simplest estimation approach is to replace the observed data quantities in the statistical estimands above with corresponding sample estimates. For example, two general approaches include parametric implementations of the g-formula and weighting \cite{robins1986, hernan2024}, which we define in the Supplement. Consistent estimates of sampling variances for the resulting estimators can be obtained using "sandwich" variance estimators in the context of M-estimation \cite{stefanski2002}, non-parametric bootstrapping \cite{efron1994introduction}, or other simulation-based methods \cite{greenland2004interval}, to account for the estimation of all model parameters. 

\section{DISCUSSION}

In this paper, we systematically organize identification strategies that can be used to combine information from an experimental study and an external population sample, specifically when the goal is to improve statistical precision or to facilitate head-to-head comparisons of treatments initially evaluated in different settings. We organized relevant insights from an extensive literature ranging from the seminal report by Pocock (1976) \cite{pocock1976combination} to more recent work with a causal orientation \cite{breskin2021fusion, ventz2022, li2023improving, wu2023comparative, van2025adaptive, valancius2024causal, shook2024fusing, karlsson2024robust, colnet2024causal}, using the language of potential outcomes to elaborate identification strategies across a family of common data structures. We propose that combining different data sources to identify causal parameters in a target population, for the purposes described herein, can be thought of as a distinct class of causal problems. Naturally, these problems inherit ideas from generalizability and transportability methods \cite{bareinboim2012transportability, westreich2017, dahabreh2019identification, dahabreh2020transportingStatMed, dahabreh2021studydesigns}, where investigators extend trial-based inferences to target populations that may differ in its distribution of effect modifiers. However, because our objective is to use external data to learn about the trial population in most applications \cite{westreich2017, rudolph2017, dahabreh2020transportingStatMed, ung2025generalizing}, differences in data structures, sampling, and identification conditions needed to construct external comparators would be obscured if these considerations were conflated with transportability analyses.

Our framework highlights the idea that the identifiability conditions needed for this work are governed by the relationships between the populations under combination, and the treatment assignment mechanisms within each population. Causal inference using these studies requires strong subject matter knowledge to apply a reasonable sampling scheme, reason about jointly intervening on trial participation and treatment, specify causal estimands, and select appropriate identification strategies with varying data requirements \cite{dahabreh2024combining}. Some relevant considerations for combining information have been previously described, including whether study engagement effects are present \cite{pocock1976combination, sacks1982, amiri2020food, freidkorn2022, fda2023}; consistency of potential outcomes and well-defined interventions within and across sources \cite{gehanfreireich1974, pocock1976combination, sacks1982, signorovitch2010comparative, thorlundextcont2020, freidkorn2022, rippinreview2022, ventz2022, valancius2024causal}, and comparability of the populations underlying the trial and external data \cite{gehanfreireich1974, pocock1976combination, doll1980, fleming1982, sacks1982, diehl1986comparison, ishak2015simulation, bareinboim2016causalfusion, amiri2020food, seeger2020, thorlundextcont2020, hall2021, li2023improving, mishra2022external, ventz2022, fda2023, signorovitch2023matching, valancius2024causal}. Our framing did provide newer newer insights, for example when pursuing identification under transportability of relative measures \cite{dahabreh2024relative, wang2024relative}; salvaging identification in the presence of study engagement effects \cite{ung2025generalizing}; and combining information from the index and external data to learn about other target populations of interest \cite{dahabreh2020toward, dahabreh2023efficient}.

The assumptions that underlie these identification strategies suggest that these endeavors should be pursued with inferential humility \cite{dahabreh2024invited}. Even in simplified settings with analytically favorable conditions and none of the complexities encountered in practical applications (e.g., loss to follow-up), drawing causal conclusions from studies that combine information rests on strong, and sometimes even heroic, assumptions. The decision to combine information across sources might invite deliberations on whether their underlying populations have similar standards of clinical care, reasonably contemporaneous, and whether data fields can be sufficiently harmonized across sources. Where possible, it is prudent to interrogate the testable implications of the appended data (e.g., under Proposition 2) using falsification methods \cite{hartman2013, lu2019causal, dahabreh2020benchmarking, dahabreh2022benchframework}, provided well-known caveats are acknowledged \cite{bancroft1944biases, giles1993pre, dahabreh2024using}. Furthermore, potential violations of the identifiability conditions can also be induced by analytic decisions made by investigators \cite{suissasat2020, chiu2022selection}. For example, recent work has shown bias may occur by conditioning on treatment within the population underlying the external data \cite{chiu2022selection}.

In sum, we have systematically organized identification strategies to learn about causal effects when combining experimental studies with external data. Future directions may include practical extensions (e.g., failure-time outcomes); causal analyses of studies that aggregate data from multiple underlying populations, e.g., in adaptive trial designs \cite{berry2015platform, adaptive2019adaptive}; and refining evidence synthesis methods \cite{dahabreh2020toward}. While researchers today have easy access to large sources of patient level data, generating valid causal inferences when combining data sources requires careful consideration of strong and largely unverifiable conditions, as well as expert judgments about whether the chosen data are sufficiently fit-for-purpose. When such conditions are met, these identification strategies may offer promise to inform patient care, medical interventions, and regulatory decision-making.

\clearpage
\renewcommand{\refname}{REFERENCES}
\printbibliography

\ddmmyyyydate 
\newtimeformat{24h60m60s}{\twodigit{\THEHOUR}.\twodigit{\THEMINUTE}.32}
\settimeformat{24h60m60s}
%

\end{document}